\documentstyle[osa,epsf,rotate,manuscript]{revtex} 
\newcommand{\dfrac}{\displaystyle\frac}

\begin{document}
%\draft
\title{\large {\bf  The pseudo-spin symmetry in Zr and Sn isotopes from the    %
proton drip line to the neutron drip line }}
\author{J. Meng$^{*,a,b}$, K.Sugawara-Tanabe$^{a,c}$ ,
S.Yamaji$^{a}$, and A.Arima$^{a}$}
\address{
$^{*}$Department of Technical Physics, Peking University,\\
Beijing 100871, P.R. China \\
$^{a}$  The Institute of Physical and Chemical Research (RIKEN)  \\
Hirosawa 2-1, Wako-shi, Saitama, 351-0198 JAPAN \\
$^{b}$Institute of Modern Physics, Chinese Academy of Sciences \\
Lanzhou 730000,  China  \\
$^{c}$ Otsuma Women's University, Tama, Tokyo 206-8540,Japan\\}
\date{\today}
\maketitle
\begin{abstract}
Based on the Relativistic continuum
Hartree-Bogoliubov (RCHB) theory, the 
pseudo-spin approximation in exotic nuclei is investigated 
in Zr and Sn isotopes from the proton drip line to the neutron drip line. 
The quality of the pseudo-spin approximation 
is shown to be connected with
the competition between the centrifugal barrier  (CB) and the
pseudo-spin orbital potential ( PSOP ). The PSOP depends on
the derivative of the difference between the scalar and 
vector potentials $dV/dr$.
If $dV/dr = 0$, the pseudo-spin symmetry is exact.
The pseudo-spin symmetry is found to be a good approximation for normal
nuclei and to become much better for exotic nuclei with
highly diffuse potential, which have $dV/dr \sim 0$.
The energy splitting of the pseudo-spin partners is smaller
for orbitals near the Fermi surface 
( even in the continuum ) than the deeply bound orbitals. 
The lower components of the Dirac wave functions for the
pseudo-spin partners are very similar and 
almost equal in magnitude.
\end{abstract}
\par
%\newpage
%\baselineskip = 24pt
%\narrowtext

%%%%%%%%%%%%%%%%%%%%%%%%%%%%%%%%%%%%
%     Introduction                 %
%%%%%%%%%%%%%%%%%%%%%%%%%%%%%%%%%%%%

\section{Introduction}

In a recent letter \cite{MSY.98}, by relating the pseudo-spin symmetry
back to the Dirac equation
through the framework of
Relativistic Continuum Hartree-Bogoliubov (RCHB)
theory\cite{ME.98}, the pseudo-spin approximation in real nuclei was discussed.
From the Dirac equation, the mechanism behind the pseudo-spin
symmetry was studied and the
pseudo-spin symmetry was shown to be related with
the competition between the centrifugal barrier  (CB) and the
pseudo-spin orbital potential ( PSOP ), which is mainly decided by
the derivative of the difference between the scalar and vector potentials.
With the  scalar and vector potentials derived from a
self-consistent RCHB calculation,
the pseudo-spin symmetry and its
energy dependence have been discussed \cite{MSY.98}.
Here we will extend our previous investigation \cite{MSY.98} to 
exotic nuclei. The pseudo-spin symmetry approximation 
for exotic nuclei is investigated
for Zr and Sn isotopes ranging from the proton 
drip line to the neutron drip line.
The isospin and energy dependence of the pseudo-spin approximation 
are investigated in detail.

The concept of pseudo-spin 
is based on the experimental observation 
that the single particle orbitals 
with $j=l+1/2$ and $j=(l+2)-1/2$ lie very close in energy and 
can therefore be labeled as pseudo-spin doublets with quantum 
number $\tilde n = n-1$, $\tilde l = l-1$, and $\tilde s = s =1/2$. 
This concept was originally found in spherical nuclei 
30 years ago \cite{AHS.69,HA.69}, 
but later proved to be a good approximation in deformed nuclei 
as well \cite{RR.73}.
It is shown that the pseudo-spin symmetry remains an important
physical concept even in the case of triaxiality \cite{BBDB.97} .

Since the suggestion of the pseudo-spin symmetry, much
efforts has been made to understand its origin. 
Apart from the rather formal relabeling of quantum numbers, various proposals
for an explicit transformation from the normal scheme to the pseudo-spin
scheme have been made in the last twenty years and several nuclear
properties have been investigated in this 
scheme \cite{Mot.91,ZM.91,BDM.92,CMQ.92,BBD.96}.
Based on the single particle Hamiltonian of the 
oscillator shell model the origin of pseudo-spin was 
proved to be connected with the special ratio in the strength 
of the spin-orbit and orbit-orbit interactions \cite{bohr,CMQ.92}
and the unitary operator performing a transformation from 
normal spin to pseudo-spin space was discussed 
\cite{CMQ.92,BBD.96,bohr,BCM.92,BCD95}.
However, it was not explained why this special
ratio is allowed in nuclei.
The relation between the pseudo-spin symmetry 
and the relativistic mean field ( RMF ) theory \cite{SW.86} was first 
noted in Ref. \cite{BDM.92}, in which Bahri et al found that the RMF 
explains approximately the strengths of spin-orbit and orbit-orbit 
interactions in the non-relativistic calculations.
In a recent paper Ginocchio took a step further and
revealed that pseudo-orbital angular
momentum is nothing but the ``orbital angular momentum'' of the lower
component of the Dirac wave function \cite{Gi97}.
He also built the
connection between the pseudo-spin
symmetry and the equality in the scalar
and vector potentials \cite{Gi97,GL.98}.

To understand to what extent it is broken in real
nuclei,  some investigation along this line has been done for square well
potentials \cite{Gi97} and for spherical solutions 
of the RMF equations \cite{GM.98}. 
By relating the pseudo-spin symmetry
back to the Dirac equation
through the framework of RCHB
theory, the pseudo-spin approximation in real nuclei 
was shown to be connected with
the competition between the centrifugal barrier  (CB) and the
pseudo-spin orbital potential ( PSOP ), which is mainly decided by
the derivative of the difference between the scalar and vector potentials.
With the  scalar and vector potentials derived from a
self-consistent Relativistic Hartree-Bogoliubov  calculation,
the pseudo-spin symmetry and its
energy dependence have been discussed in Ref. \cite{MSY.98}.

The highly unstable nuclei with extreme proton and neutron
ratio are now accessible with the help of the radioactive nuclear beam
facilities. The physics connected with the extreme neutron richness in 
these nuclei and the low density in the tails of their distributions  have
attracted more and more attention not only
in nuclear physics but also in other fields such as 
astrophysics \cite{TA.95,HJJ.95}.
New exciting discoveries have been made
by exploring hitherto inaccessible regions in the nuclear
chart.
It is very interesting to investigate the pseudo-spin symmetry 
approximation both in normal and exotic nuclei. For this purpose, we will 
use the RCHB 
theory, which is the extension of the
RMF and the Bogoliubov transformation
in the coordinate representation, and provides not only a unified
description of the mean field and pairing correlation 
but also the proper description for the continuum and its 
coupling with the bound state \cite{ME.98,MR.96}.
As this theory takes into account the proper isospin dependence
of the spin-orbit term,
it is able to provide a good description of global
experimental data not only for stable nuclei but also for 
exotic nuclei throughout the nuclear chart \cite{ME.98}. 
It is very interesting to examine the pseudo-spin symmetry 
approximation in exotic nuclei, in which the mean 
field potentials are expected to be highly diffuse.

Recently, by relating the pseudo-spin symmetry 
back to the Dirac equation through the framework of RCHB
theory, the pseudo-spin approximation in real nuclei was discussed.
The mechanism behind the pseudo-spin
symmetry was studied and the
pseudo-spin symmetry was shown to be connected with
the competition between the CB and the PSOP, which is mainly decided by
the derivative of the difference between the scalar and vector potentials.
With the  scalar and vector potentials derived from a
self-consistent RHB calculation,
the pseudo-spin symmetry and its
energy dependence have been discussed \cite{MSY.98}.
Here we will extend the previous investigation to the 
case of exotic nuclei.
The pseudo-spin splitting in Zr and Sn isotopes  has been studied
from the proton drip line to the neutron drip line.
The energy splitting of the pseudo-spin partners, their  
energy  and isospin dependence will be addressed.
An outline of the RCHB formalism is briefly reviewed
in Sec. II.
In Sec. III, the Dirac equation and the formalism leading 
to the pseudo-spin symmetry is presented. 
The  energy splitting of the pseudo-spin partners and
its energy dependence are given in Sec. IV.
The  pseudo-spin orbital potential, which breaks the 
pseudo-spin symmetry will be studied in Sec. V. 
In Sec. VI, the wave-function of pseudo-spin partners will 
be studied. 
A brief summary is given in the last section.

%%%%%%%%%%%%%%%%%%%%%%%%%%%%%%%%%%%%
%     FORMALISM                    %
%%%%%%%%%%%%%%%%%%%%%%%%%%%%%%%%%%%%

\section{An outline of RCHB Theory}   

The RCHB theory is obtained by combining the 
RMF and the Bogoliubov transformation
in the coordinate representation \cite{MR.96}, and its 
detailed formalism and numerical
solution can be found in Ref. \cite{ME.98} and the references therein.
The RCHB theory can give a fully self-consistent description of the chain
of Lithium isotopes \cite{MR.96} ranging
from $^{6}$Li to $^{11}$Li. The  halo in $^{11}$Li has
been successfully reproduced in this
self-consistent picture and excellent agreement with recent
experimental data is obtained.  
The contribution from the continuum has been taken into account 
and proved to be crucial to understand the halo in exotic nuclei. 
Based on the RCHB, a new phenomenon "Giant Halo"
has been predicted. The "Giant Halo" is
composed not only of one or two neutrons, as is the
case in the halos in light $p$-shell
nuclei, but also up to 6 neutrons \cite{MR.97}.
The development of skins and halos and their relation with the 
shell structure are systematically studied
with RCHB in Ref. \cite{MTY.97}, where both the pairing and blocking
effect have been treated self-consistently. Therefore the 
RCHB theory is very suitable for the examination 
of the pseudo-spin approximation in exotic nuclei.

The basic ansatz of the RMF theory starts from a Lagrangian
density by which nucleons are described
as Dirac particles interacting via the exchange of various
mesons and the photons.
The mesons considered are the scalar
sigma ($\sigma$), vector omega ($\bf \omega$) and iso-vector vector rho
($\bf \vec \rho$).  The iso-vector vector rho
($\bf \vec \rho$) meson provides the necessary isospin
asymmetry.
The scalar sigma meson moves in the self-interacting field of cubic and
quadratic terms with strengths $g_2$ and $g_3$, respectively.
The Lagrangian then consists of the free baryon
and meson parts and the interaction part with minimal coupling,
together with the nucleon mass $M$, and $m_\sigma$,
$g_\sigma$, $m_\omega$, $g_\omega$, $m_\rho$, $g_\rho$  the masses and
coupling constants of the respective mesons:
\begin{equation}
\begin{array}{rl}
{\cal L} &= \bar \psi (i\rlap{/}\partial -M) \psi +
        \,{1\over2}\partial_\mu\sigma\partial^\mu\sigma-U(\sigma)
        -{1\over4}\Omega_{\mu\nu}\Omega^{\mu\nu}\\
        \ &+ {1\over2}m_\omega^2\omega_\mu\omega^\mu
        -{1\over4}{\vec R}_{\mu\nu}{\vec R}^{\mu\nu} +
        {1\over2}m_{\rho}^{2} \vec\rho_\mu\vec\rho^\mu
        -{1\over4}F_{\mu\nu}F^{\mu\nu} \\
        & -  g_{\sigma}\bar\psi \sigma \psi~
        -~g_{\omega}\bar\psi \rlap{/}\omega \psi~
        -~g_{\rho}  \bar\psi
        \rlap{/}\vec\rho
        \vec\tau \psi
        -~e \bar\psi \rlap{/}A \psi.
\label{Lagrangian}
\end{array}
\end{equation}

The field tensors for the
vector mesons are given as:
\begin{eqnarray}
\left\{
\begin{array}{lll}
   \Omega^{\mu\nu}   &=& \partial^\mu\omega^\nu-\partial^\nu\omega^\mu, \\
   {\vec R}^{\mu\nu} &=& \partial^\mu{\vec \rho}^\nu
                        -\partial^\nu{\vec \rho}^\mu
                        - g^{\rho} ( {\vec \rho}^\mu 
                           \times {\vec \rho}^\nu ), \\
   F^{\mu\nu}        &=& \partial^\mu \vec A^\nu-\partial^\nu \vec A^\mu.
\end{array}   \right.
\label{tensors}
\end{eqnarray}
For a realistic description of
nuclear properties,  a nonlinear self-coupling of the scalar
mesons turns out to be crucial \cite{BB.77}:
\begin{equation}
   U(\sigma)~=~\dfrac{1}{2} m^2_\sigma \sigma^2_{}
            ~+~\dfrac{g_2}{3}\sigma^3_{}~+~\dfrac{g_3}{4}\sigma^4_{}
\end{equation}

The classical variation principle gives the following equations of
motion :
\begin{equation}
   [ {\vec \alpha} \cdot {\vec p} +
     V_V ( {\vec r} ) + \beta ( M + V_S ( {\vec r} ) ) ]
     \psi_i ~=~ \epsilon_i\psi_i
\label{spinor1}
\end{equation}
for the nucleon spinors and
\begin{eqnarray}
\left\{
\begin{array}{lll}
  \left( -\Delta \sigma ~+~U'(\sigma) \right ) &=& -g_\sigma\rho_s
\\
   \left( -\Delta~+~m_\omega^2\right )\omega^{\mu} &=&
                    g_\omega j^{\mu} ( {\vec r} )
\\
   \left( -\Delta~+~m_\rho^2\right) {\vec \rho}^{\mu}&=&
                    g_\rho \vec j^{\mu}( {\vec r} )
\\
          -\Delta~ A_0^{\mu} ( {\vec r} ) ~ &=&
                           e j_{\rho}^{\mu}( {\vec r} )
\end{array}  \right.
\label{mesonmotion}
\end{eqnarray}
with $U'(\sigma) = \partial_\sigma U(\sigma)$ and 
$\Delta = - \partial^\mu\partial_\mu$ for the mesons, where
\begin{eqnarray}
\left\{
\begin{array}{lll}
   V_V ( {\vec r} ) &=&
      g_\omega\rlap{/}\omega + g_\rho\rlap{/}\vec\rho\vec\tau
         + \dfrac{1}{2}e(1-\tau_3)\rlap{\,/}\vec A , \\
   V_S ( {\vec r} ) &=&
      g_\sigma \sigma( {\vec r} ) \\
\end{array}
\right.
\label{vaspot}
\end{eqnarray}
are the vector and scalar potentials respectively and
the source terms for the mesons are
\begin{eqnarray}
\left\{
\begin{array}{lll}
   \rho_s &=& \sum_{i=1}^A \bar\psi_i \psi_i
\\
   j^{\mu} ( {\vec r} ) &=&
               \sum_{i=1}^A \bar \psi_i \gamma^{\mu} \psi_i
\\
   \vec j^{\mu}( {\vec r} ) &=&
          \sum_{i=1}^A \bar \psi_i \gamma^{\mu} \vec \tau  \psi_i
\\
   j^{\mu}_p ( {\vec r} ) &=&
      \sum_{i=1}^A \bar \psi_i \gamma^{\mu} \dfrac {1 - \tau_3} 2  \psi_i,
\end{array}  \right.
\label{mesonsource}
\end{eqnarray}
where the summations are over the valence nucleons only.
It should be noted that as usual, the present approach
neglects the contribution of
negative energy states, i.e.,  no-sea approximation,
which means that the vacuum
is not polarized. The coupled equations Eq.(\ref{spinor1})
and Eq.(\ref{mesonmotion}) are nonlinear quantum
field equations, and their exact solutions are very complicated.
Thus the mean field approximation is generally used: i.e.,
the meson field operators in Eq.(\ref{spinor1}) are replaced
by their expectation values, so that the nucleons move
independently in the classical meson fields. The
coupled equations are self-consistently solved by iteration.
\par

For spherical nuclei, i.e., the systems with rotational symmetry,
the potential of the nucleon and the sources of
meson fields depend only on the radial coordinate $r$. The spinor
is characterized by the quantum numbers $l$, $j$,$m$, and the
isospin $t = \pm \dfrac 1 2$ for neutron and proton, respectively.
The other quantum number is denoted by  $i$. 
The Dirac spinor has the form:
\begin{equation}
   \psi ( \vec r ) = \left( g \atop f \right) =
      \left( { {\mbox{i}  \dfrac {G_i^{lj}(r)} r {Y^l _{jm} (\theta,\phi)} }
      \atop
       { \dfrac {F_i^{lj}(r)} r (\vec\sigma \cdot \hat {\vec r} )
       {Y^l _{jm} (\theta,\phi)} } }
      \right) \chi _{t}(t),
\label{reppsi}
\end{equation}
where $Y^l _{jm} (\theta,\phi)$ are the spinor spherical harmonics
and $G_i^{lj}(r)$ and $F_i^{lj}(r)$ are the radial
wave function for upper and lower components. They are normalized
according to the relation:
\begin{equation}
 \int_0^{\infty}dr ( | G_i^{lj}(r) |^2 + | F_i^{lj}(r) |^2 ) = 1.
\end{equation}
The radial equation of spinor Eq. (\ref{spinor1}) can be reduced as :
\begin{eqnarray}
\left\{
\begin{array}{lll}
   \epsilon_i G_i^{lj}(r) &=& ( - \dfrac {\partial} {\partial r}
      + \dfrac {\kappa_i} r )  F_i^{lj}(r) + ( M + V_S(r) + V_V(r) ) G_i^{lj}(r)
\\
   \epsilon_i F_i^{lj}(r) &=& ( + \dfrac {\partial} {\partial r}
      + \dfrac {\kappa_i} r )  G_i^{lj}(r)
      - ( M + V_S(r) - V_V(r) ) F_i^{lj}(r) ,
\end{array}  \right.
\label{spinorradical}
\end{eqnarray}
where 
\begin{displaymath}
   \kappa =
      \left\{
         \begin{array}{ll}
            -(j+1/2)  & for ~ j=l+1/2 \\
            +(j+1/2)  & for ~ j=l-1/2. \\
         \end{array}
      \right.
\end{displaymath}
The meson field equations become simply radial Laplace equations
of the form:
\begin{equation}
   \left( - \frac {\partial^2} {\partial r^2}  - \frac 2 r
   \frac  {\partial}   {\partial r} + m_{\phi}^2 \right)\phi = s_{\phi} (r),
\label{Ramesonmotion}
\end{equation}
$m_{\phi}$ are the meson masses for $\phi = \sigma, \omega,\rho$
and for photon ( $m_{\phi} = 0$ ). The source terms are:
\begin{eqnarray}
   s_{\phi} (r) = \left\{
   \begin{array}{ll}
   -g_\sigma\rho_s - g_2 \sigma^2(r)  - g_3 \sigma^3(r)
   & { \rm for ~ the ~  \sigma~  field } \\
   g_\omega \rho_v    & {\rm for ~ the ~ \omega ~ field} \\
   g_{\rho}  \rho_3(r)       & {\rm for~ the~ \rho~ field} \\
   e \rho_c(r)  & {\rm for~ the~ Coulomb~ field}, \\
\end{array}
\right.
\end{eqnarray}

\begin{eqnarray}
\left\{
\begin{array}{lll}
   4\pi r^2 \rho_s (r) &=& \sum_{i=1}^A ( |G_i(r)|^2 - |F_i(r)|^2 ) \\
   4\pi r^2 \rho_v (r) &=& \sum_{i=1}^A ( |G_i(r)|^2 + |F_i(r)|^2 ) \\
   4\pi r^2 \rho_3 (r) &=& \sum_{p=1}^Z ( |G_p(r)|^2 + |F_p(r)|^2 )
               -  \sum_{n=1}^N ( |G_n(r)|^2 + |F_n(r)|^2 ) \\
   4\pi r^2 \rho_c (r) &=& \sum_{p=1}^Z ( |G_p(r)|^2 + |F_p(r)|^2 ) . \\
\end{array}
\right.
\label{mesonsourceS}
\end{eqnarray}
The Laplace equation can be solved by using the Green function:
\begin{equation}
\phi (r) = \int_0^{\infty} r'^2 dr' G_{\phi} (r,r') s_{\phi}(r'),
\end{equation}
where for massive fields
\begin{equation}
G_{\phi} (r,r') = \frac 1 {2m_{\phi}} \frac 1 {rr'}
( e^{-m_{\phi} | r-r'|} -  e^{-m_{\phi} | r+r'|} )
\end{equation}
and for Coulomb field
\begin{equation}
G_{\phi} (r,r') =
\left\{
\begin{array}{ll}
1/r  & {\rm for ~ r~ >~ r' } \\
1/r' & {\rm for ~r~ <~ r'} .\\
\end{array}
\right.
\end{equation}

The Eqs.(\ref{spinorradical}) and (\ref{Ramesonmotion}) could be solved 
self-consistently in the usual RMF approximation.  
However, Eq.(\ref{spinorradical}) does not contain
the pairing interaction, as the classical meson fields 
are used in RMF. In order to have the 
pairing interaction, one has to 
quantize the meson fields which leads to a Hamiltonian with two-body 
interaction. Following the standard procedure of 
Bogoliubov transformation, a Dirac Hartree-Bogoliubov equation 
could be derived and then a unified description of the mean 
field and pairing correlation in nuclei 
could be achieved. For the details, see 
Ref. \cite{ME.98} and the references therein. 
The RHB equations are as following:
\begin{equation}
   \int d^3r'
   \left( \begin{array}{cc}
          h-\lambda &   \Delta \\
          \Delta    &  - h+\lambda
          \end{array} \right) 
   \left( { \psi_U \atop\psi_V } \right)  ~
   = ~ E ~ \left( { \psi_U \atop \psi_V } \right), 
\label{ghfb}
\end{equation}
where 
\begin{equation}
   h(\vec r,\vec r') =  \left[ {\vec \alpha} \cdot {\vec p} +
      V_V( {\vec r} ) + \beta ( M + V_S ( {\vec r} ) ) \right]
      \delta(\vec r,\vec r')
\label{NHamiltonian}
\end{equation}
is the Dirac Hamiltonian and the Fock term has been neglected as is 
usually done in RMF. The pairing potential is :
\begin{eqnarray}
    \Delta_{kk'}(\vec r, \vec r')
    &=& - \int d^3r_1 \int d^3r_1' \sum_{\tilde k \tilde k'}
    V_{kk',\tilde k \tilde k'} ( \vec r \vec r'; \vec r_1 \vec r_1' )
    \kappa_{\tilde k \tilde k'}  (\vec r_1, \vec r_1' ).
\label{gap}
\end{eqnarray}
It is obtained from the one-meson exchange interaction
$V_{kk',\tilde k \tilde k'} (\vec r \vec r'; \vec r_1 \vec r_1' )$ in the
$pp$-channel and the pairing tensor $\kappa=V^*U^T$: 
\begin{eqnarray}
   \kappa_{k k'}( \vec r, \vec r') = < | a_{k} a _{k'} | >
   =  \psi_V^{k}(\vec r) ^* \psi_U^{k'}(\vec r)^T.
\end{eqnarray}
The nuclear density is as following:
\begin{eqnarray}
   \rho(\vec r ,\vec r' ) 
   = \sum_{ilj} g_{ilj} \psi_V^{ilj} (\vec r) ^* \psi_V ^{ilj} (\vec r ' ) .
\end{eqnarray}
As in Ref. \cite{ME.98}, $V$ used for the 
pairing potential in Eq.(\ref{gap}) is either the density-dependent 
two-body force of zero range 
with the interaction strength $V_0$ and the nuclear matter density $\rho_0$:
\begin{equation}
   V(\mbox{\boldmath $r$}_1,\mbox{\boldmath $r$}_2) = V_0
     \delta(\mbox{\boldmath $r$}_1-\mbox{\boldmath $r$}_2)
     \frac{1}{4}\left[1-
     \mbox{\boldmath $\sigma$}_1\mbox{\boldmath $\sigma$}_2\right]
     \left(1 - \frac{\rho(r)}{\rho_0}\right),
\label{vpp}
\end{equation}
or Gogny-type finite range force 
with the parameter $\mu_i$, $W_i$, $B_i$, $H_i$ and
$M_i$ ($i=1,2$): \cite{BGG.84} 
\begin{equation}
   V(\mbox{\boldmath $r$}_1,\mbox{\boldmath $r$}_2)
      ~=~\sum_{i=1,2}
       e^{((\mbox{\boldmath $r$}_1-\mbox{\boldmath$r$}_2) / \mu_i)^2}
       (W_i + B_i P^{\sigma} - H_i P^{\tau} - M_i P^{\sigma} P^{\tau}).
\label{vpp2}
\end{equation}
A Lagrange multiplier $\lambda$ is introduced to fix the 
particle number for the neutron
and proton as $N = \mbox {Tr} \rho _n$ and  $Z = \mbox {Tr} \rho _p$ .

In order to describe both continuum and bound states 
self-consistently, we use the RHB theory in  
coordinate representation, i.e., the
Relativistic Continuum Hartree-Bogolyubov ( RCHB )
theory \cite{ME.98}. It is then applicable to both 
exotic nuclei and normal nuclei.
In Eq. (\ref{ghfb}), the
eigenstates occur in pairs of opposite energies. 
When spherical symmetry is imposed on the solution of the
RCHB equations, the wave function can be 
written as:
\begin{equation}
   \psi^i_U = \left( {\displaystyle {\mbox{i}  \frac {G_U^{ilj}(r)} r }  \atop
     {\displaystyle \frac {F_U^{ilj}(r)} r 
        (\vec\sigma \cdot \hat {\vec r} )  } }
              \right) {Y^l _{jm} (\theta,\phi)}  \chi_{t}(t) ,  
   \psi^i_V =
       \left( {\displaystyle {\mbox{i} \frac {G_V^{ilj}(r)} r }  \atop
          {\displaystyle \frac {F_V^{ilj}(r)} r 
             (\vec\sigma \cdot \hat {\vec r} )
       } } \right)  {Y^l _{jm} (\theta,\phi)}  \chi_{t}(t).
\end{equation}

Using the above equation, Eq.(\ref{ghfb})  depends only 
on the radial coordinates and can be
expressed as the following integro-differential equation:
\begin{eqnarray}
\left\{
   \begin{array}{lll}
      \displaystyle
      \frac {d G_U(r)} {dr} + \frac {\kappa} r G_U(r) -
       ( E + \lambda-V_V(r) + V_S(r) ) F_U(r) +
         r \int r'dr' \Delta(r,r')  F_V(r') &=& 0  \\
      \displaystyle
      \frac {d F_U(r)} {dr} - \frac {\kappa} r F_U(r) +
       ( E + \lambda-V_V(r)-V_S(r) ) G_U(r) +
         r \int r'dr' \Delta(r,r') G_V(r') &=& 0 \\
      \displaystyle
      \frac {d G_V(r)} {dr} + \frac {\kappa} r G_V(r) +
       ( E - \lambda+V_V(r)-V_S(r) ) F_V(r) +
         r \int r'dr' \Delta(r,r') F_U(r') &=& 0 \\
      \displaystyle
      \frac {d F_V(r)} {dr} - \frac {\kappa} r F_V(r) -
       ( E - \lambda+V_V(r)+V_S(r) ) G_V(r) +
         r \int r'dr' \Delta(r,r')  G_U(r') &=& 0, \\
\end{array}
\right.
\label{CoupEq}
\end{eqnarray}
where the nucleon mass is included in the 
scalar potential $V_S(r)$.
For the $\delta$-force of Eq.(\ref{vpp}), Eq.(\ref{CoupEq})
is reduced to normal coupled differential equations and can be solved
with shooting method by Runge-Kutta algorithms.
For the case of Gogny force, the coupled integro-differential equations 
are discretized in the space and solved by the 
finite element methods. The numerical details 
can be found in Ref. \cite{ME.98}. Now we have to solve 
Eqs.({\ref{CoupEq}) and (\ref{Ramesonmotion})
self-consistently for the RCHB case. 
As the calculation with Gogny force is very time-consuming,
we solve them only for one case in order to fix the interaction
strength for $\delta$-force in Eq.(\ref{vpp}).

\section{The pseudo-spin symmetry}

The Dirac equation in RMF or in the canonical basis of RCHB 
describes a Dirac spinor with mass $M$ moving in a 
scalar potential $V_S( {\vec r} )$ and a vector 
potential $V_V( {\vec r} )$.
With $\epsilon=M+E$, the 
potential $V = V_V( {\vec r} ) + V_S( {\vec r} )$, which 
is around $-50$ MeV, and the effective mass $M^*=M+V_S( {\vec r} )$,
the relation between the upper and lower 
components of the wave function can be written as:
\begin{equation}
   \left\{ \begin{array}{ccc}
      g &=& \displaystyle\frac 1 {E - V} ( {\vec {\sigma} \cdot {\vec p}} )~~ f 
           \nonumber\\  \nonumber\\
      f &=& \displaystyle\frac 1 {E + 2M^* - V} 
         ( {\vec {\sigma} \cdot {\vec p}}  ) ~~g
   \end{array} \right. 
\label{spinor2}
\end{equation}
Then the coupled equations are reduced  to uncoupled ones 
for the upper and lower components, respectively.
Effectively we get the corresponding 
Schr\"odinger equation for both components:
\begin{equation}
   \left\{ \begin{array}{cc}
     ( \vec {\sigma} \cdot {\vec p} )
       \displaystyle\frac {1} {E +2 M^* - V} 
        ( {\vec {\sigma} \cdot {\vec p}} )  g
     =  ( E - V )  g \nonumber\\  
       ( \vec {\sigma} \cdot {\vec p} )
       \displaystyle\frac {1} {E - V} 
       ( {\vec {\sigma} \cdot {\vec p}} )  f
     =  ( E + 2 M^* -V ) f
   \end{array} \right. 
\label{smaspinor}
\end{equation}
In the spherical case, $V$ depends only on the radius.  
We choose the phase convention of the vector spherical harmonics 
as:
\begin{eqnarray} 
( \vec \sigma \cdot \vec r ) Y^l_{jm} = - Y^{l'}_{jm}
\end{eqnarray}
where 
\begin{eqnarray}
       l' = 2j-l=
        \left\{ \begin{array}{cc}
            l+1,  &  j=l+1/2  \\
            l-1,  &  j=l-1/2
         \end{array} \right.
\end{eqnarray}
Here $l'$ is nothing but the pseudo-orbital angular momentum $\tilde l$. 
After some tedious procedures, one gets the radial equation 
for the lower and upper components respectively:
\begin{eqnarray}
\small
   &&   [ \frac {d^2} {dr^2} 
     + \frac 1 {E - V} \frac {dV} {dr} \frac {d} {dr}] F^{lj}_i(r)
   \nonumber\\ 
   &&  +  [ \frac { \kappa ( 1 - \kappa ) }  {r^2} 
     - \frac 1 {E - V} \frac {\kappa} r \frac {dV} {dr}] F^{lj}_i(r) 
\nonumber\\
 = &-&  ( E + 2M^* - V )  ( E - V ) F^{lj}_i(r),
\label{smaspinor4}
\end{eqnarray}
\begin{eqnarray}
   & &  [ \frac {d^2} {dr^2}  - \frac 1  {E + 2M^* - V} 
        \frac {d(2M^* - V)} {dr} \frac {d} {dr} ] G^{lj}_i(r) \nonumber \\
   &-&  [  \frac { \kappa ( 1 + \kappa ) }  {r^2} 
        + \frac 1 {E + 2M^* - V} \frac {\kappa} r 
        \frac {d(2M^* - V)} {dr}  ] G^{lj}_i (r) \nonumber \\
 = &-&  (E + 2M^*- V ) ( E - V ) G^{lj}_i (r), 
\label{larspinor4} 
\end{eqnarray}
where 
\begin{eqnarray}
   \kappa ( \kappa - 1 ) = l'(l'+1),  ~~~~~ \kappa ( \kappa + 1 ) =l(l+1).
\end{eqnarray}
It is clear that one can use either Eq.(\ref{smaspinor4}) or 
equivalently Eq.(\ref{larspinor4}) to get the eigenvalues $E$ and
the corresponding eigenfunctions.
Normally Eq.(\ref{larspinor4}) is used in the literature and
the spin-orbital splitting is discussed in connection with
the corresponding  spin-orbital potential
$\displaystyle \frac 1 {E + 2M^* - V} \frac {\kappa} r
        \frac {d(2M^* - V)} {dr}$.
If the Eq. (\ref{smaspinor4}) is used instead and the 
pseudo spin-orbital potential ( PSOP ) term,
$\displaystyle\frac 1  {E - V}  \frac {\kappa} r \frac {dV} {dr}$,
is neglected, then the eigenvalues $E$ for the same $l'$ will 
degenerate. This is the phenomenon of pseudo-spin symmetry 
observed in \cite{AHS.69,HA.69}. It means that Eq.(\ref{spinor2})
is the transformation between the normal spin formalism and the 
pseudo-spin formalism. 

In Eq. (\ref{smaspinor4}), the term which splits the pseudo-spin 
partners is simply the PSOP. 
The hidden symmetry for the pseudo-spin approximation 
is revealed as $dV/dr = 0$, which is more 
general and includes $V = 0$ discussed in \cite{Gi97} as a special 
case. For exotic nuclei with highly diffuse potentials, 
$dV/dr \sim 0$ may be a good approximation and then the pseudo-spin 
symmetry will be good.
But generally, $dV/dr = 0$ is not always satisfied 
in the nuclei and the pseudo-spin symmetry is an 
approximation.  However, 
if $| \displaystyle\frac 1  {E - V}  \frac {\kappa} r 
\frac {dV} {dr}| \ll |\displaystyle\frac { \kappa ( 1 - \kappa ) }  {r^2}| $, 
the pseudo-spin approximation will be good. 
Thus, the comparison of the relative magnitude of the 
centrifugal barrier ( CB ), 
$\displaystyle\frac { \kappa ( 1 - \kappa ) }  {r^2} $,  
and the PSOP can provide us with some information on the  
pseudo-spin symmetry. 

In a recent letter \cite{MSY.98}, 
the mechanism behind the pseudo-spin
symmetry was studied and the
pseudo-spin symmetry was shown to be connected with
the competition between the centrifugal barrier  (CB) and the
pseudo-spin orbital potential ( PSOP ), which is mainly decided by
the derivative of the difference between the scalar and vector potentials.
With the  scalar and vector potentials derived from a
self-consistent RCHB calculation,
the pseudo-spin symmetry and its
energy dependence have been discussed.
Here in this paper we will extend the previous investigation to the
case of exotic nuclei. The pseudo-spin symmetry 
for exotic nuclei is investigated
for Zr and Sn isotopes from the proton drip line to the neutron drip line.
The isospin and energy dependence of the pseudo-spin approximation
will be investigated in detail in the following section.

\section{The energy splitting of the pseudo-spin partners}

We use here the non-linear Lagrangian parameter set NLSH \cite{SNR.93}
which could provide a good 
description of all nuclei from oxygen to lead. As we 
study not only the closed shell nuclei, but also the open 
shell nuclei, the inclusion of the pairing is necessary.
The pairing interaction strength is the same as in Ref. \cite{MR.97}. 
The interaction strength in the pairing force of zero range
Eq.(\ref{vpp}) is properly renormalized by  
the calculation of RCHB with Gogny force. 
Since we use a pairing force of zero range,
we have to limit the number of continuum levels
by a cut-off energy.  For each spin-parity channel, 20
radial wave functions are taken into account, which
corresponds roughly to a cut-off energy of
120 MeV for a fixed box radius $R=20$ fm. 
For the fixed cut-off energy and for the box radius
$R$, the strength $V_0$ of the pairing force in Eq.(\ref{vpp}) 
is determined by adjusting the corresponding pairing energy
$-\frac{1}{2}\mbox{Tr}\Delta\kappa$ to that of a
RCHB-calculation using the finite range part of the Gogny
force D1S \cite{BGG.84}. We use the nuclear
matter density 0.152 fm$^{-3}$ for $\rho_0$.

The quality of pseudo-spin symmetry can be understood more clearly by
considering the microscopic structure of the wave
functions and the single particle energies in the canonical
basis. As shown in Ref. \cite{ME.98}, the particle levels 
for the bound states in canonical basis are the same 
as those by solving the Dirac equation with the scalar and
vector potentials from RCHB. 
Therefore Eqs. (\ref{smaspinor4}) and (\ref{larspinor4}) 
are valid in canonical basis after the
pairing interaction has been taken into account and 
are very suitable for the discussion of the pseudo-spin 
symmetry.

The neutron single particle levels in
$^{150}$Sn and $^{120}$Zr are given in 
Fig. 1a and 1b, respectively.
The four sets of pseudo-spin partners, i.e.,
$1d_{3/2}$ and $2s_{1/2}$,
$1f_{5/2}$ and $2p_{3/2}$, $1g_{7/2}$ and $2d_{5/2}$,
$2d_{3/2}$ and $3s_{1/2}$, are marked by boxes. 
As seen in the figure, the energy splitting between pseudo-spin partners
decreases with the decreasing binding energy.
The single particle energy of $3s_{1/2}$ in $^{120}Zr$
is $-6.00$ MeV, and its
partner  $2d_{3/2}$ is $-5.86$ MeV, the splitting is $0.14$ MeV.
While $2s_{1/2}$ is $-31.62$ MeV,  $1d_{3/2}$ $-33.23$ Mev, the
splitting is $1.61$ MeV, which is bigger than the former one
by a factor of $10$. The same situation is found for 
the energy splitting between pseudo-spin partners in $^{150}$Sn: 
The single particle energies of $3s_{1/2}$  and $2d_{3/2}$ 
are $-9.645$ and $-10.11$ MeV, respectively. 
The single particle energies for other 
pseudo-spin partners in $^{150}$Sn are,  
$-11.74$ and $-13.87$ MeV for  $2d_{5/2}$ and $1g_{7/2}$ partners, 
$-22.46$ and $-25.50$ MeV for 
$2p_{3/2}$ and $1f_{5/2}$ partners, 
$-33.63$ and $-36.58$ for $2s_{1/2}$ and $1d_{3/2}$ partners, respectively. 
Although we show only the  neutron single particle levels in
$^{150}$Sn and $^{120}$Zr as examples here, the same are 
found in other Sn and Zr isotopes. It is usually seen that 
the pseudo-spin symmetry approximation becomes better near the Fermi surface,
which is in agreement with the experimental observation. 

In Fig.1 there are also two pairs of pseudo-spin partners 
( $3p_{3/2}$ and $2f_{5/2}$ partners and 
$2f_{7/2}$ and $1h_{9/2}$ partners ) near the threshold, apart from 
the fours pairs of pseudo-spin partners below the Fermi level. 
The energies for these two pairs of pseudo-spin partners are 
$-1.581$ and $-1.031$ Mev for the $3p_{3/2}$ and $2f_{5/2}$ partners, and
$-2.549$ and $-2.620$ MeV for the $2f_{7/2}$ and $1h_{9/2}$ partners, 
respectively. 
Considering their pseudo-spin orbital angular momentum 
$\tilde l = 2$ and $4$, their splittings  
$\Delta E = \displaystyle \frac {E_{\tilde lj=\tilde l-1/2}-
    E_{\tilde lj=\tilde l+1/2}} {2\tilde l+1}$  
are only $-0.1100$ MeV and $0.7889 \times 10^{-2}$ MeV, respectively. 
This is due to the energy dependence and the diffuseness of the 
potential in exotic nuclei, which we will discuss in the following. 
As it is seen in Fig.1, the normal splitting is such that 
the orbital $j = \tilde l + 1/2$ is 
below the orbital $j = \tilde l - 1/2$, except for 
$3p_{3/2}$ and $2f_{5/2}$ partners. 
The same also happens for $2d_{3/2}$ and $3s_{1/2}$ 
partners in Zr isotopes. 
The pseudo-spin splitting depends on the  
derivative of the difference between the scalar
and vector potentials $dV/dr$, which is small 
for the exotic nuclei with
highly diffuse potential. The integration of 
$\dfrac {dV} {dr} |F|^2$ over $r$ 
gives the splitting of the pseudo-spin partners, 
whose sign will decide the normal splitting or the reverse.
The subtle details 
of the potential are crucial for the  pseudo-spin splitting.

To see the behavior of the pseudo-spin partners around the 
Fermi level and the isospin dependence of the pseudo-spin splitting, 
we show the single particle levels near 
the Fermi surface in the canonical basis for the Sn and Zr 
isotopes with an even neutron number as a function of the mass
number in Fig. 2. The Fermi level is shown by the dashed line. 
The pseudo-spin splitting for the 
pseudo-spin partners,  
$2d_{3/2}$ and $3s_{1/2}$,  remains small 
in Zr and Sn isotopes from the proton drip line to the neutron drip line.
The pseudo-spin symmetry 
remains even valid for exotic nuclei. 
The pseudo-spin symmetry near the neutron drip line 
becomes better than that near the $\beta$-stability line. In Fig. 2a, 
there is a kink for the single
particle levels in the continuum, as the contribution 
from the continuum becomes important and the potential becomes 
diffuse around $^{130}$Sn. But 
the splitting for $3p_{3/2}$ and $2f_{5/2}$ partners,  and
$2f_{7/2}$ and $1h_{9/2}$ partners 
in Sn isotopes is small and the
pseudo-spin symmetry approximation is very good, 
independent of whether they are 
in the continuum or near the threshold.
Therefore we can see that the pseudo-spin symmetry 
is very well reserved for the orbital near the threshod energy and 
in the continuum region.

In order to see the energy dependence and the isospin dependence 
of the pseudo-spin orbital splitting 
more clearly, we plot 
$\Delta E = \displaystyle \frac {E_{\tilde lj=\tilde l-1/2}-
    E_{\tilde lj=\tilde l+1/2}} {2\tilde l+1}$ 
versus
$E = \displaystyle \frac
     {  \tilde l E_{\tilde lj=\tilde l + 1/2 } 
     + ( \tilde l + 1 ) E_{ \tilde lj= \tilde l-1/2} } {2 \tilde l+1} $
for the bound pseudo-spin partners in Sn and Zr isotopes in Fig. 3. 
In both isotopes, a monotonous
decreasing  behavior with the decreasing binding energy 
is clearly seen. The pseudo-spin splitting
for $3s_{1/2}$ and $2d_{3/2}$ is more than $10$ times smaller
than that of the $2s_{1/2}$ and $1d_{3/2}$. 
As far as the isospin dependence
of the pseudo-spin orbital splitting is concerned, 
the splitting in Sn isotopes gives a monotonous
decreasing  behavior with the increasing isospin. Particularly 
for $2s_{1/2}$ and $1d_{3/2}$ partners, the pseudo-spin splitting 
in $^{170}$Sn is only half of that in $^{96}$Sn. Just as 
we expected, the pseudo-spin symmetry in neutron-rich nuclei is better.
In Zr isotopes, although the situation is more 
complicated ( e.g., the effect of the deformation which is 
neglected here ) , 
the pattern is more or less the same, i.e., a monotonous
decreasing  behavior with the decreasing binding energy 
and a monotonous decreasing  behavior with the isospin. 
From these studies, we see that the 
pseudo-spin symmetry remains a good approximation for
both stable and exotic nuclei.
A better pseudo-spin symmetry  can 
be expected for the orbital near the threshold, 
particularly for nuclei near the particle drip-line.

\section{The pseudo-spin orbital potential}

To understand why the energy splitting of the pseudo-spin 
partner changes with different binding energies and why the 
pseudo-spin approximation is good in RMF, the PSOP 
and CB should be examined carefully. Unfortunately, 
it is very hard to compare them clearly, as 
the PSOP has a singularity at $E \sim V$. 
As we are only interested in the relative magnitude of the
CB and the PSOP, we introduce the effective CB,
$\displaystyle (E-V) \frac {\kappa(\kappa-1)} {r^2}$,
and the effective PSOP,
$\displaystyle\frac {\kappa} {r} \frac {dV} {dr}$,
for comparison.
They correspond to the CB and the PSOP
multiplied by a common factor $E-V$ respectively.

The effective PSOP
does not depend on the binding energy of the single particle level,
but depends on the angular momentum and parity.
On the other hand the effective CB
depends on the energy. Comparing these two
effective potentials one could see
the energy dependence of the pseudo-spin symmetry.
They are given in Fig. 4 
for $s_{1/2}$ ( lower ) and  $d_{3/2}$ ( upper )
of $^{120}$Zr in arbitrary scale. 

The pseudo-spin approximation is much better
for the less bound pseudo-spin partners, because the effective CB
is smaller for the more deeply bound states.
This is in agreement with the
results shown in Fig.3.
The effective PSOP and the effective CB
are also given as inserts in Fig.4 in order to show their
behavior near the nuclear surface. 

In order to examine this carefully,
we compare the effective CB
( dashed lines or dot-dashed lines ) and  the
effective PSOP ( solid lines ) multiplied
by the squares of the lower
component wave function $F(r)$, which are given in Fig. 5,
for $2s_{1/2}$ ( upper left ), $3s_{1/2}$ ( lower left),
$1d_{3/2}$ ( upper right ), and  $2d_{3/2}$ ( lower right )
of $^{120}$Zr in arbitrary scale.
The pseudo-spin approximation is much better
for the less bound pseudo-spin partners, because the effective CB
is smaller for the more deeply bound states.
This is in agreement with the
results shown above.
The integrated values of the potentials in Fig.2 with $r$ are
proportional to their contribution to the energy
after some proper renormalization. It is
clear that the contribution of the effective CB
( dashed lines or dot-dashed lines ) is much bigger than that of the
effective PSOP ( solid lines ). Generally
the effective PSOP is two orders 
of magnitude smaller than the effective CB. 

In Fig. 1 and 2 we notice that
the orbital $j = \tilde l + 1/2$ is generally 
below the orbital $j = \tilde l - 1/2$, except for
$3p_{3/2}$ and $2f_{5/2}$ partners in Sn isotopes.
The same situation also happens for $2d_{3/2}$ and $3s_{1/2}$
partners in Zr isotopes. As the pseudo-spin splitting depends 
on PSOP, which depends on the subtle radial dependence of the 
potentials, sometimes the PSOP may have positive or 
negative regions as a function of $r$ which cancell each other. 
	The integration of $\dfrac {dV} {dr} |F|^2$ 
	over $r$ gives the splitting of the pseudo-spin partners,
	whose sign will decide the normal splitting or the reverse.
	That is the reason why 
	the orbital $j = \tilde l + 1/2$ is above 
	the orbital $j = \tilde l - 1/2$  for
	$3p_{3/2}$ and $2f_{5/2}$ partners in Sn isotopes and 
	for $2d_{3/2}$ and $3s_{1/2}$ partners in Zr isotopes.

	\section{The wave-function of pseudo-spin partners}

	In the above discussion, we have seen that the PSOP 
	is much smaller than the CB. Therefore if we neglect the PSOP 
	in Eq.(\ref{smaspinor4}), the lower component of the 
	Dirac wave functions for the
	pseudo-spin partners will be the same,
	i.e., in the case of the exact pseudo-spin symmetry,
	the lower component of the pseudo-spin partners should be identical (except
	for the phase).
	The upper component of the
	Dirac wave functions can be obtained from the transformation 
	in Eq.(\ref{spinor2}), which depends on 
	the  quantum number $\kappa$. Therefore the study of 
	the Dirac wave functions for the
	pseudo-spin partners will provide a check for the 
	pseudo-spin approximation in nuclei. 
	As examples, the normalized single nucleon wave functions  for 
	the upper ( $G$ ) and lower ( $F$ )
	components of the Dirac wave functions for the
	pseudo-spin partners $1d_{3/2}$ and $2s_{1/2}$,
	$1f_{5/2}$ and $2p_{3/2}$, $1g_{7/2}$ and $2d_{5/2}$,
	$2d_{3/2}$ and $3s_{1/2}$ in $^{120}$Zr are given in Fig. 6. 
	Of course, the lower components are much smaller in magnitude compared
	with the upper component in Eq.(\ref{spinor2}).
	The phase of the Dirac wave functions for one of
	the pseudo-spin partners has been reversed in order to
	have a careful comparison. It is seen that the 
	lower components of the
	Dirac wave functions for the
	pseudo-spin partners are very similar and are 
	almost equal in magnitude, as observed also for 
	$^{208}$Pb in Ref. \cite{GM.98}. 
	The similarity in the lower components $F$ of the wave function 
	for the pseudo-spin partners near the 
	Fermi surface is better than for the deeply bound ones.
	The lower components for the pseudo-spin partners 
	with small pseudo-spin orbital angular momentum are 
	better than for the ones with large 
	pseudo-spin orbital angular momentum.
	As seen in Fig.6, the similarity for pseudo-spin partners 
	$2d_{3/2}$ and $3s_{1/2}$ is better than for pseudo-spin 
	partners $1d_{3/2}$ and $2s_{1/2}$. 
	The similarities for  pseudo-spin partners, 
	$2d_{3/2}$ and $3s_{1/2}$, $1d_{3/2}$ and $2s_{1/2}$,
	are better than for the pseudo-spin partners
	$1f_{5/2}$ and $2p_{3/2}$, $1g_{7/2}$ and $2d_{5/2}$.
	 
	Although the lower components for the pseudo-spin partners are
	very close to each other, the difference for the upper 
	components is very big. The upper component of the
	Dirac wave functions can be obtained from the transformation
	in Eq.(\ref{spinor2}), which for the sperical case can be 
	reduced to the follows:
	\begin{eqnarray}
	   \displaystyle
	       G^{lj}_i (r)  =  \frac 1 {E - V} 
	       [ - \frac {dF^{lj}_i(r)} {dr}
	     +  \frac { \kappa }  {r} F^{lj}_i(r) ].
	\end{eqnarray}

	As seen in Fig.6, in the case of exact pseudo-spin symmetry, where 
	both $E$ and $F^{lj}_i(r)$ are identical for the pseudo-spin 
	partners, the upper 
	conpnonents $G^{lj}_i (r)$ will be different due to 
	the term $\displaystyle\frac { \kappa }  {r} F^{lj}_i(r)$. 
	For the pseudo-spin partners with small $\tilde l$, 
	the contribution of the term 
	$\displaystyle\frac { \kappa }  {r} F^{lj}_i(r)$ 
	becomes less important for larger $r$ and a similarity between the
	upper components can happen in the nuclear surface. As for examples, 
	for $r \geq 6$ fm,  the upper components for the
	pseudo-spin partners $1d_{3/2}$ and $2s_{1/2}$,
	$1f_{5/2}$ and $2p_{3/2}$, $2d_{3/2}$ and $3s_{1/2}$, 
	in Fig. 6 are very similar.

	\section{Summary}

	In conclusion, the pseudo-spin symmetry is examined
	in normal and exotic nuclei in the framework of
	RCHB theory. Based on RCHB theory the
	pseudo-spin approximation in exotic nuclei is investigated
	in Zr and Sn isotopes from the proton drip line to the neutron drip line.
	The quality of the pseudo-spin approximation
	is shown to be connected with
	the competition between the centrifugal barrier  (CB) and the
	pseudo-spin orbital potential ( PSOP ), which is mainly decided by
	the derivative of the difference between the scalar and
	vector potentials $dV/dr$.
	If the derivative of the difference between the scalar
	and vector potentials $dV/dr$
	vanishes, the pseudo-spin symmetry is exact.
	The condition $dV/dr \sim 0$  may be a good approximation 
	for the exotic nuclei with highly diffuse potential.
	Further the new condition 
	$\displaystyle\frac 1  {E - V}  \frac {\kappa} r
	\frac {dV} {dr} \ll \displaystyle\frac { \kappa ( 1 - \kappa ) }  {r^2} $
	is found under which the symmetry is preserved 
	approximately. We have examined this condition to see 
	how good the pseudo-spin symmetry is in RCHB.
	For a given angular momentum and parity channel,
	the effective CB, $\displaystyle (E-V) \frac {\kappa(\kappa-1)} {r^2}$, 
	becomes stronger for the less bound level, so
	the pseudo-spin symmetry for the weakly bound state is
	better than that for the deeply bound state, which 
	is in agreement with the experimental
	observation \cite{AHS.69,HA.69}.
	The pseudo-spin symmetry is found to be a good approximation even
	for the exotic nuclei with highly diffuse potential.
	The above conclusion has been well supported
	by RCHB calculations for Zr and Sn isotopes from the proton 
	drip line to the neutron drip line.
	From the simple Dirac equation, it has been shown that
	there are two equivalent ways to solve  
	the coupled Dirac equation for the upper and lower components:
	i.e., the normal spin formalism and pseudo-spin formalism.
	Both formalisms are equivalent as far as the 
	energies and wave functions are concerned.
	Their relation is given by Eq.(\ref{spinor2}),
	which indicates that the unitary transformation from the 
	conventional formalism to the 
	pseudo-spin formalism has the "p-helicity"\cite{BCD95,BBD.96,GL.98}.
	Summarizing our investigation, we conclude:

	\begin{itemize}
	\item[1]  The quality of the pseudo-spin approximation
	is connected with the competition between the CB, and the
	PSOP which is mainly proportional to 
	the derivative of the difference between the scalar and
	vector potentials $dV/dr$;

	\item[2]  The pseudo-spin symmetry is a good approximation for normal 
	nuclei and become much better for exotic nuclei with
	highly diffuse potentials;

	\item[3]  The pseudo-spin symmetry has strong energy dependence. 
	The energy splitting between the pseudo-spin partners is smaller
	for orbitals near the Fermi surface.

	\item[4]  The energy difference between the orbital $j = \tilde l + 1/2$ 
	and the orbital $j = \tilde l - 1/2$ is always negative, except for
	$3p_{3/2}$ and $2f_{5/2}$ partners.
	The same situation also happens for $2d_{3/2}$ and $3s_{1/2}$
	partners in Zr isotopes. The integration of 
	$\dfrac {dV} {dr} |F|^2$ 
over $r$ gives the splitting of the pseudo-spin partners,
whose sign will decide the normal splitting or the reverse.

\item[5] The lower components of the Dirac wave functions for the
pseudo-spin partners are very similar and
almost equal in magnitude.
The similarity in the lower components  of 
the wave function for the pseudo-spin partners near the
Fermi surface is closer than for the deeply bound ones.

\end{itemize}

%We would like to express our gratitude to T.Wright for
%his careful reading of the manuscript.

%%%%%%%%%%%%%%%%%%%%%%%%%%%%%%%%%%%%
%     References                   %
%%%%%%%%%%%%%%%%%%%%%%%%%%%%%%%%%%%%

\newpage

\leftline{\Large {\bf Figure Captions}}
\parindent = 2 true cm
\begin{description}

\item[Fig. 1] The  single particle levels
in the canonical basis for the neutron in
$^{120}$Zr and $^{150}$Sn. The Fermi surface
is shown by a dashed line. The bound pseudo-spin partners 
are marked by boxes

\item[Fig. 2] The single particle energies of the neutron 
in the canonical basis as a function of the mass number
for Zr and Sn isotopes. The dashed line indicates the
chemical potential.

\item[Fig. 3] The pseudo-spin orbit splitting 
$\Delta E = \displaystyle \frac {E_{\tilde lj=\tilde l-1/2}-
    E_{\tilde lj=\tilde l+1/2}} {2\tilde l+1}$
versus the binding energy 
$E = \displaystyle \frac
     {  \tilde l E_{\tilde lj=\tilde l + 1/2 }
     + ( \tilde l + 1 ) E_{ \tilde lj= \tilde l-1/2} } {2 \tilde l+1} $
for Zr and Sn isotopes. 
From left to 
right, the pseudo-spin partners correspond to
($1d_{3/2}, 2s_{1/2}$), ($1f_{5/2}, 2p_{3/2}$),
($1g_{7/2}, 2d_{5/2}$) and ($2d_{3/2}, 3s_{1/2}$), respectively. 

\item[Fig. 4] The comparison of the effective centrifugal barrier ( CB )
$\displaystyle (E-V) \frac {\kappa(\kappa-1)} {r^2}$
( dashed lines and dot-dashed lines )
and the effective pseudo-spin orbital potential ( PSOP )
$\displaystyle \frac {\kappa} {r} \frac {dV} {dr}$
( solid line ) in arbitrary
scale for $d_{3/2}$ ( upper ) and $s_{1/2}$ ( lower )
in $^{120}$Zr. The dashed lines are for $1d_{3/2}$ and
$2s_{1/2}$, and the dot-dashed lines are for $2d_{3/2}$ and
$3s_{1/2}$. The inserted boxes show
the same quantities, but the ordinate is magnified and the
abscissa is reduced to
show the behaviors of the effective CB
and the effective PSOP near the nuclear surface.

\item[Fig. 5] The comparison of the effective centrifugal barrier ( CB )
$\displaystyle (E-V) \frac {\kappa(\kappa-1)} {r^2}$
( dashed lines and dot-dashed lines )
and the effective pseudo-spin orbital potential ( PSOP )
$\displaystyle \frac {\kappa} {r} \frac {dV} {dr}$
( solid line ) multiplied by the square of the
wave function $F$ of the lower components in arbitrary
scale for $d_{3/2}$ ( upper ) and $s_{1/2}$ ( lower )
in $^{120}$Zr. The dashed lines are for $1d_{3/2}$ and
$2s_{1/2}$, and the dot-dashed lines are for $2d_{3/2}$ and
$3s_{1/2}$. 

\item[Fig. 6] The upper component $G$ and lower component $F$ 
of the Dirac wave functions for the 
pseudo-spin partners in $^{120}$Zr. The phase of 
the Dirac wave functions for one of 
the pseudo-spin partners has been reversed in order to 
have a careful comparison. 

\end{description}
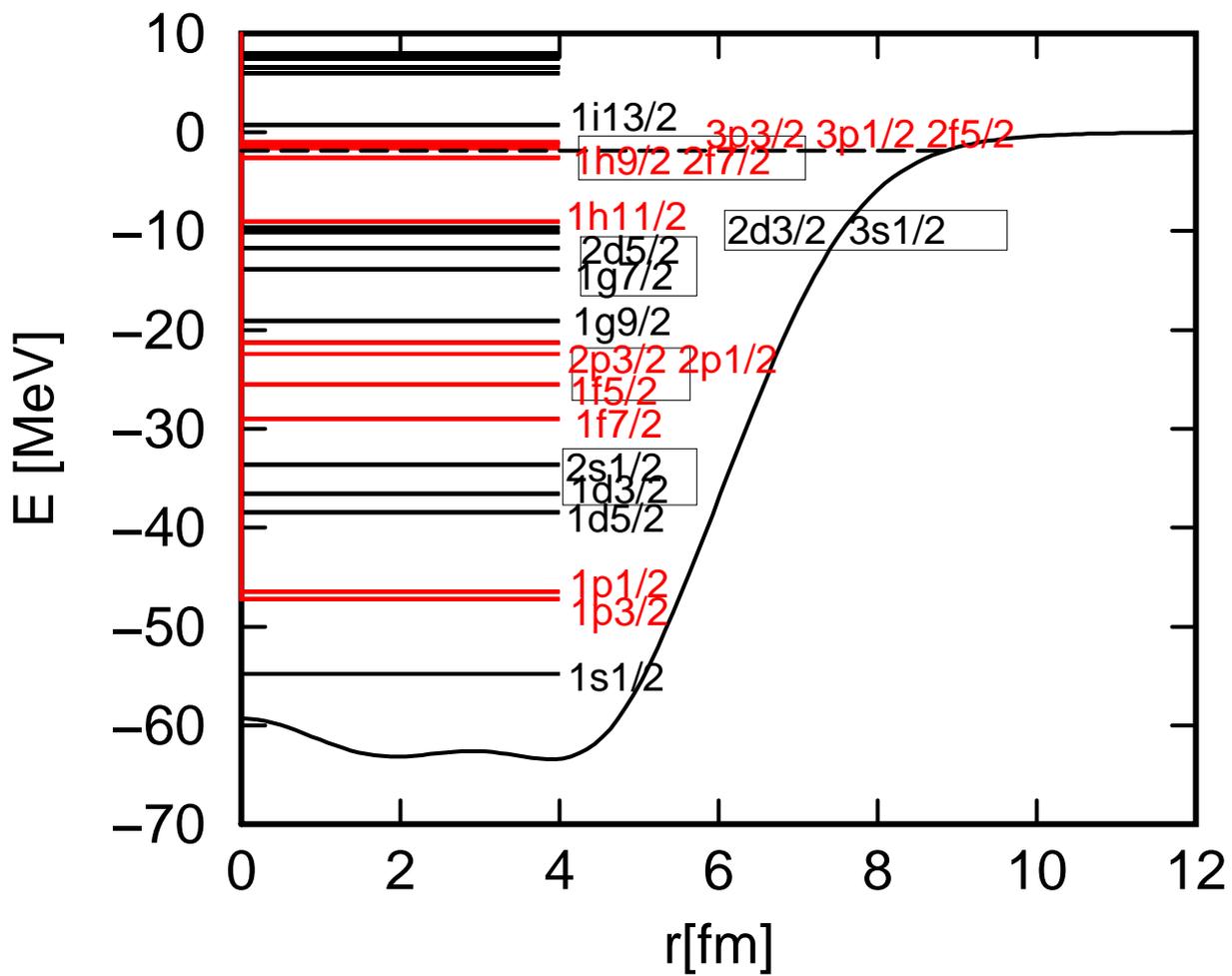
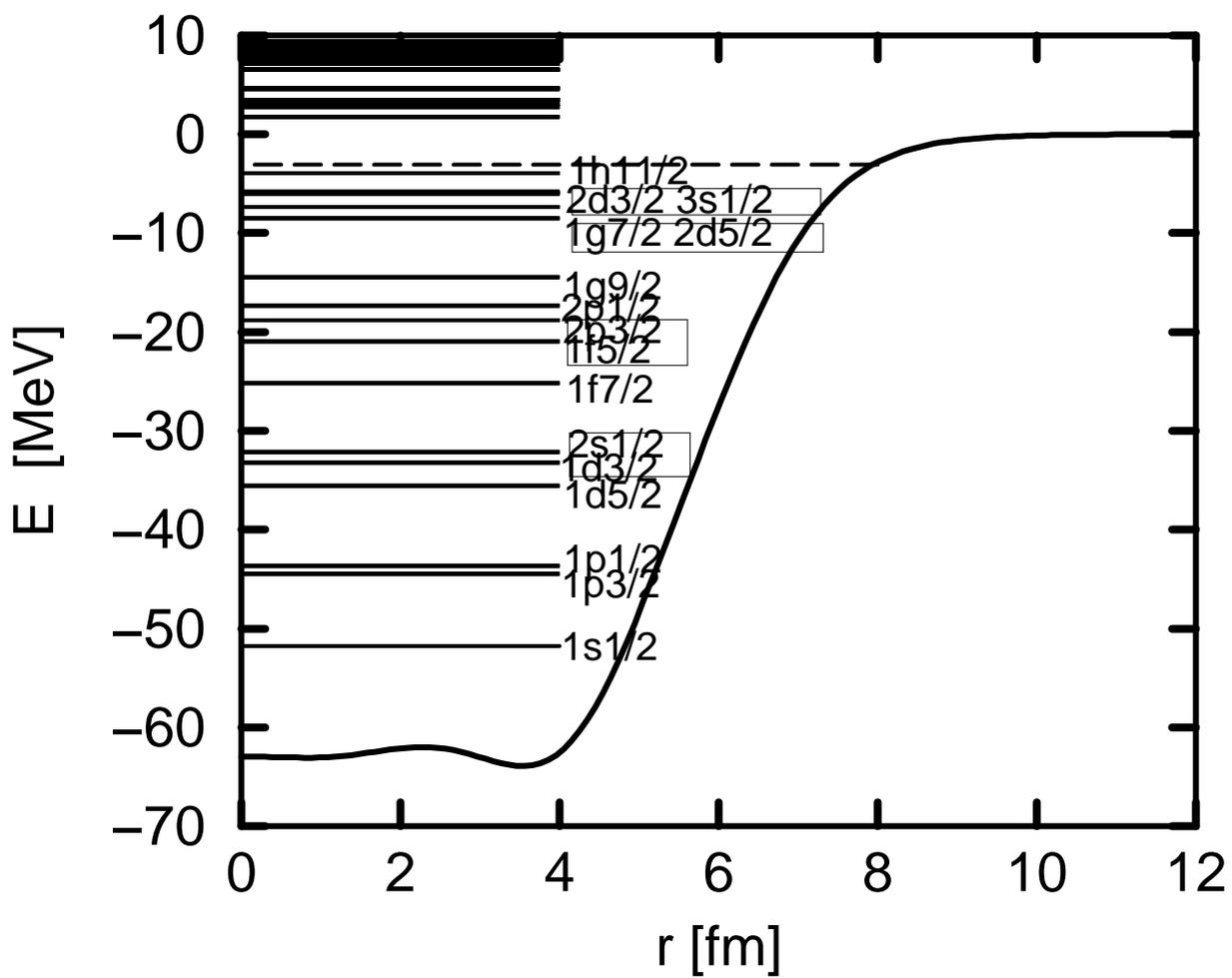
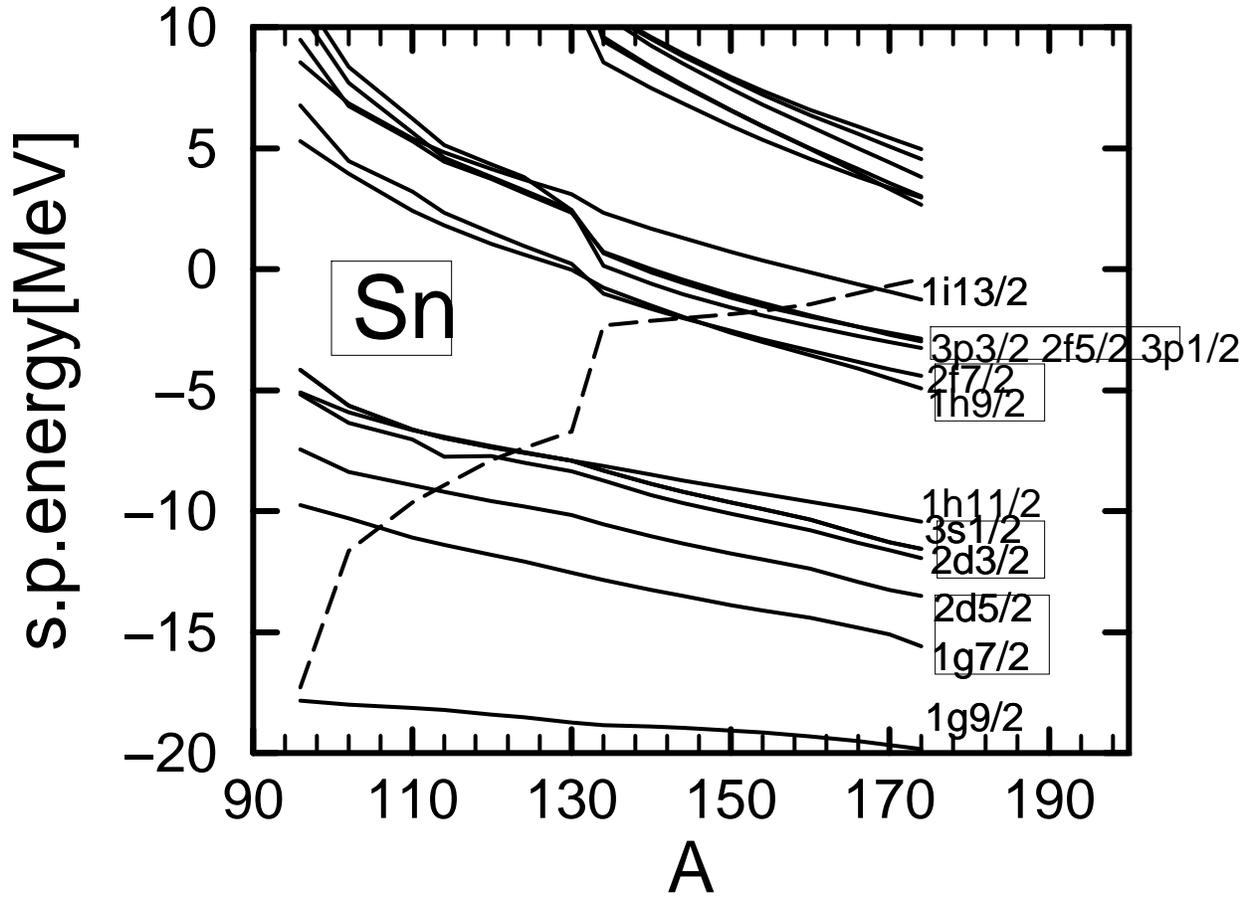
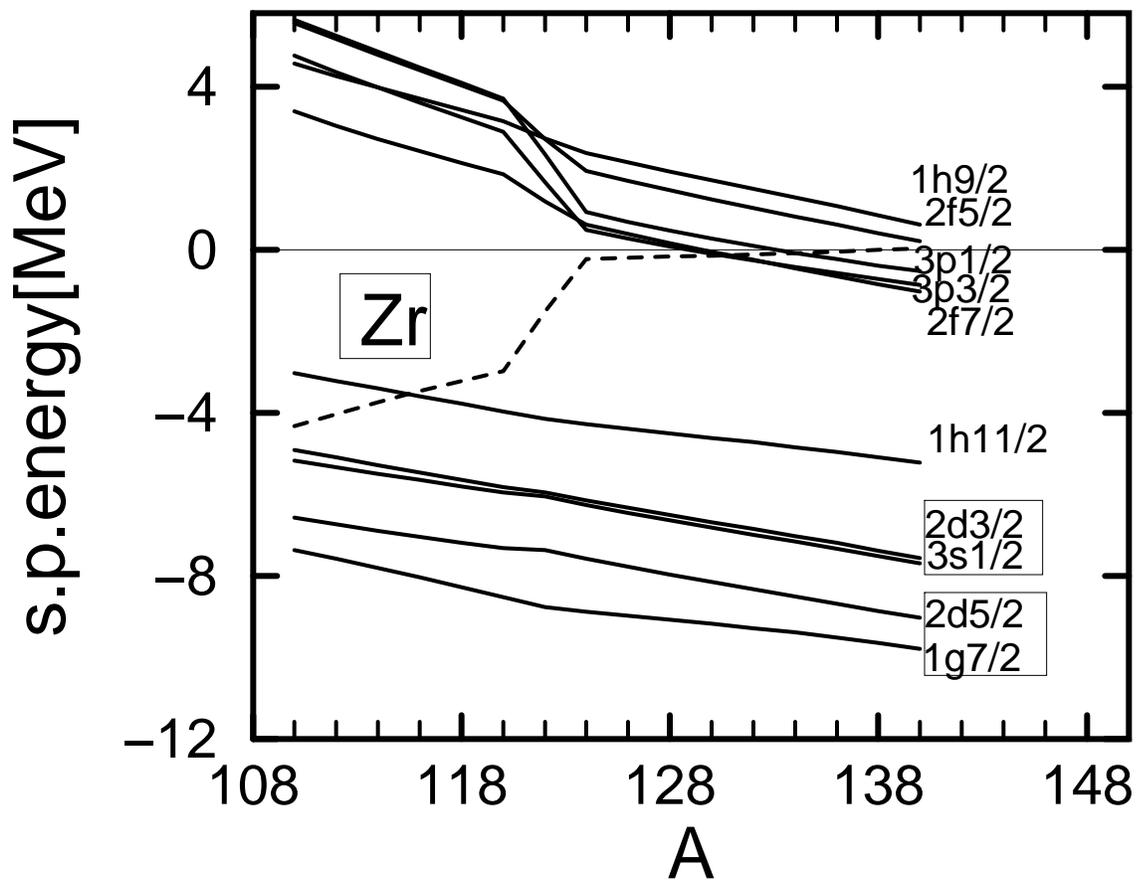
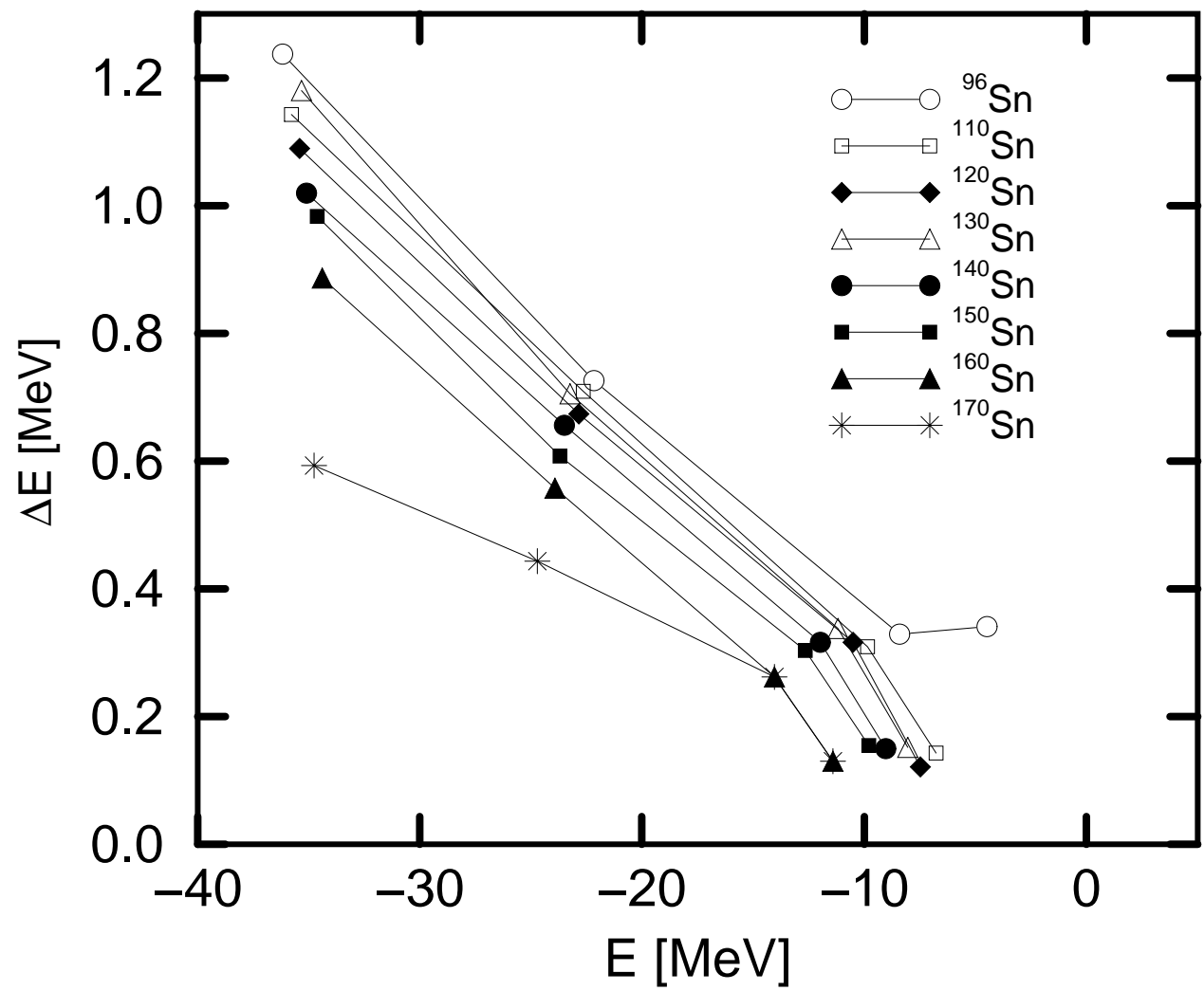
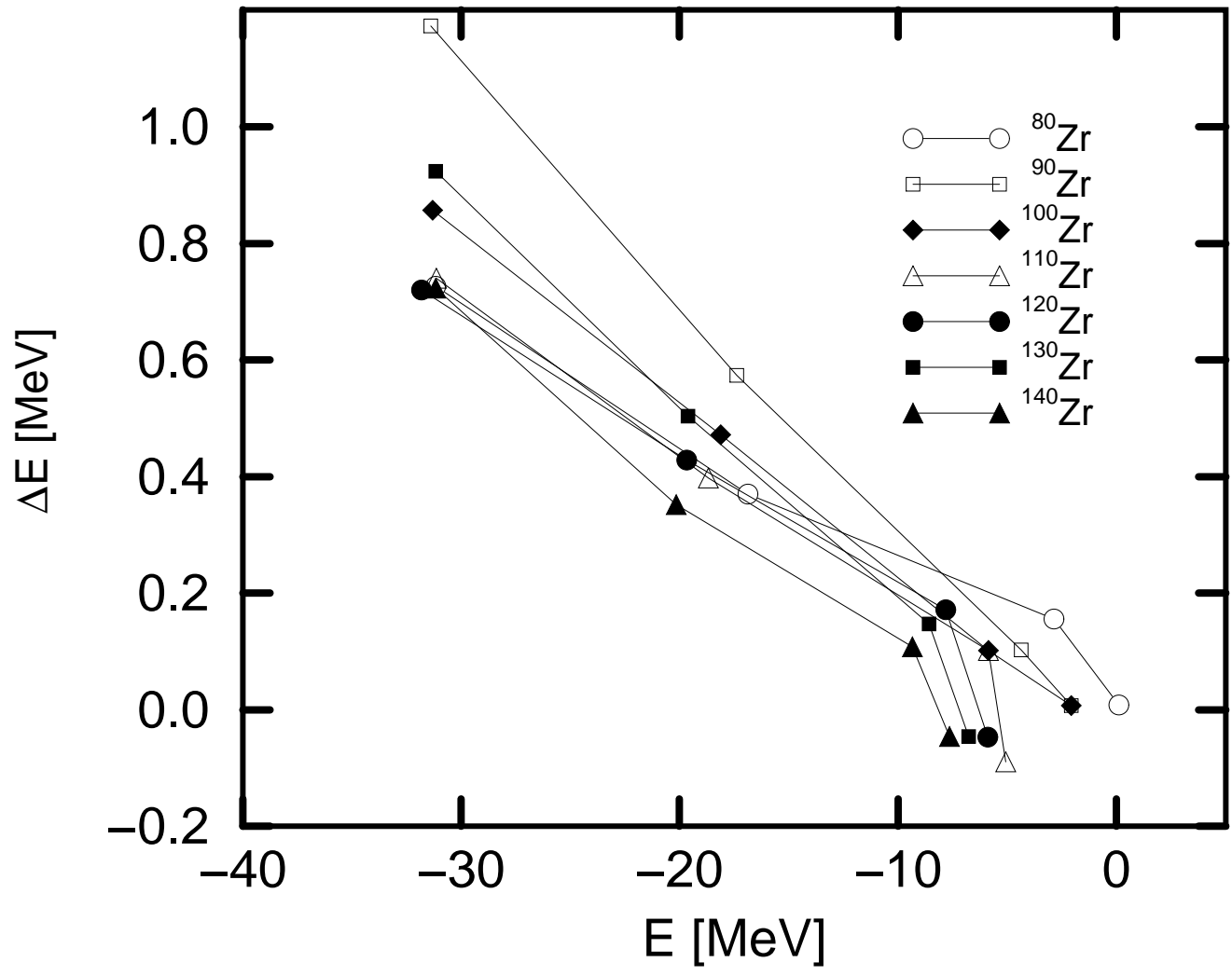
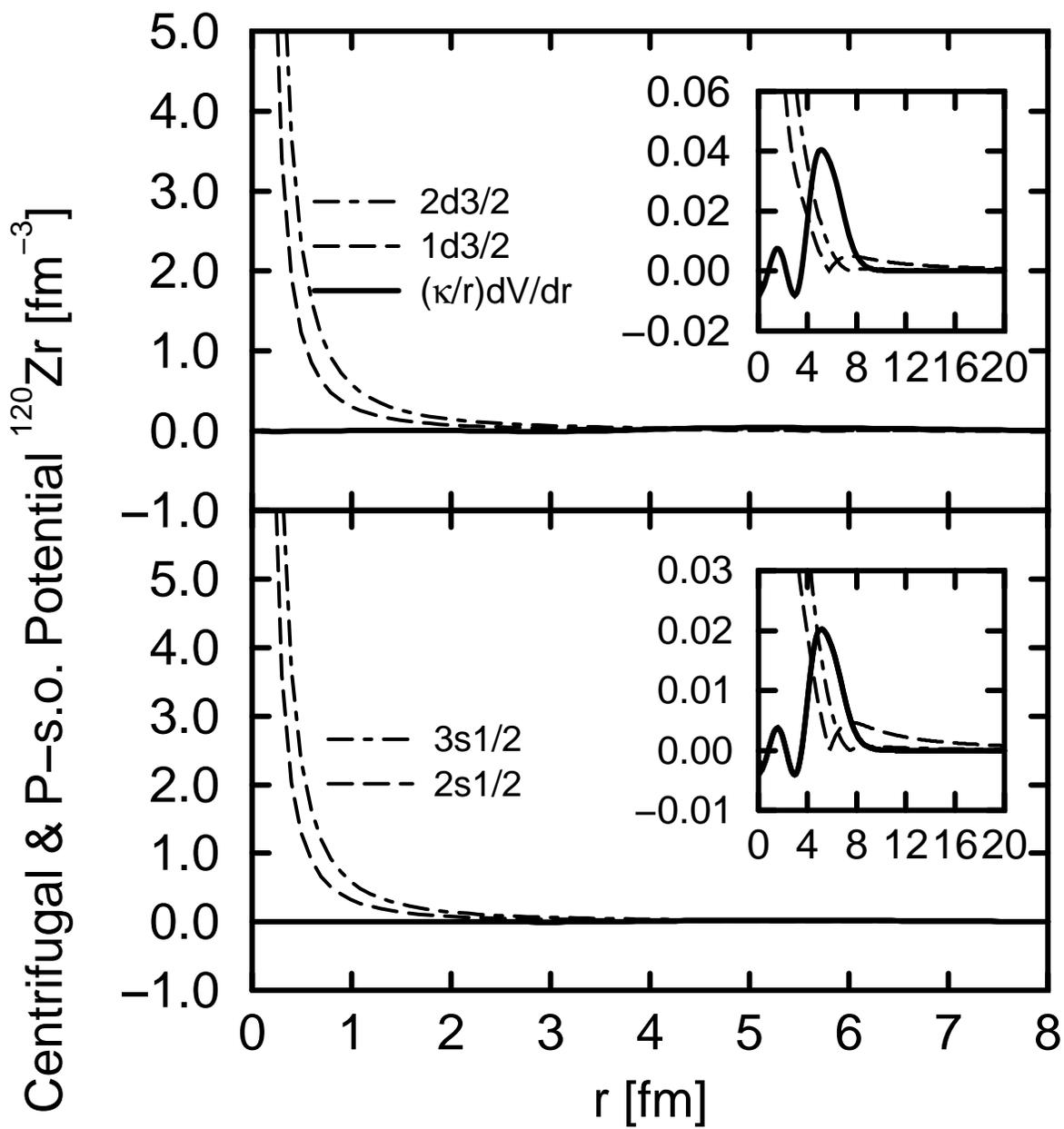
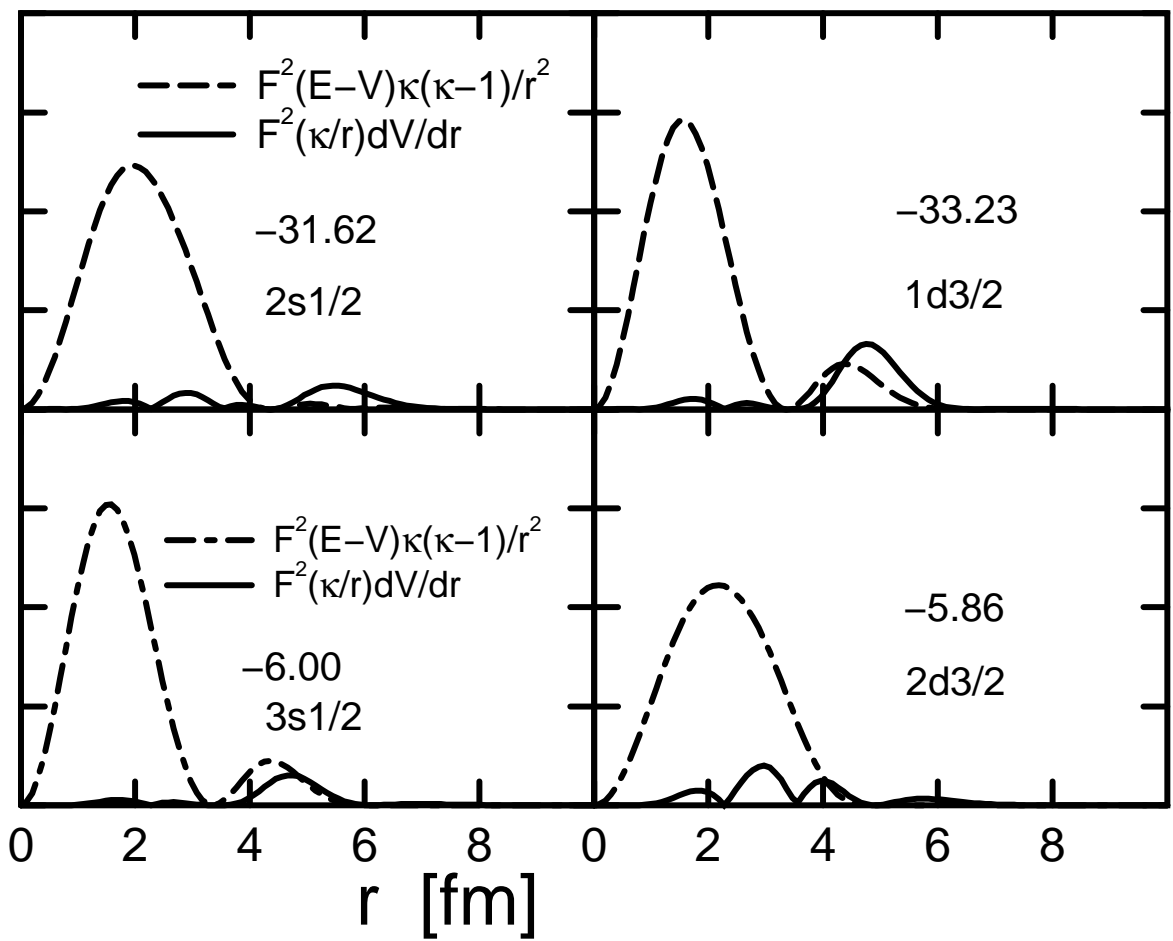
\end{document}